\documentclass[twocolumn,aps,showpacs,pre]{revtex4}
\usepackage[T1]{fontenc}
\usepackage[dvips]{graphicx}
\usepackage{amsmath}

\begin{document}
\title{A framework to investigate the immittance responses for finite length-situations: fractional diffusion equation, reaction term, and boundary conditions}
\author{E. K. Lenzi$^1$, M. K. Lenzi$^2$, F. R. G. B. Silva$^1$, G. Gon\c{c}alves$^3$, R. Rossato$^4$, R. S. Zola$^4$, and L. R. Evangelista$^1$}
\affiliation{$^1$Departamento de F\'{i}sica, Universidade Estadual de Maring\'a, Avenida Colombo 5790, 87020-900 Maring\'a, Paran\'a, Brazil.\\ $^2$ Departamento de Engenharia Qu\'{\i}mica,
Universidade Federal do Paran\'a, Setor de Tecnologia -- Jardim das Américas, Caixa Postal 19011, 81531-990, Curitiba -- PR, Brazil.\\ $^3$Departamento de Engenharia Qu\'{\i}mica,
Universidade Estadual de Maring\'a, Avenida Colombo 5790, 87020-900 Maring\'a, Paran\'a, Brazil.\\
$^4$ Universidade Tecnológica Federal do Paraná - Câmpus Apucarana, Rua Marcílio Dias 635, 86812-460 Apucarana, Paran\'a, Brazil}
\date{\today}

\begin{abstract}
The Poisson-Nernst-Planck (PNP) diffusional model for the immittance or impedance spectroscopy response of an electrolytic cell in a finite-length situation is extended to a general framework. In this new formalism, the bulk behavior of the mobile charges is governed by a fractional diffusion equation  in the presence of a reaction term. The solutions have to satisfy a general boundary condition embodying, in a single expression, most of the surface effects commonly encountered in experimental situations. Among these effects, we specifically consider the charge transfer process from an electrolytic cell to the external circuit and the adsorption-desorption phenomenon at the interfaces. The equations are exactly solved in the small AC signal approximation and are used to obtain an exact expression for the electrical impedance as a funcion of the frequency.  The predictions of the model are compared to and found to be in good agreement with the experimental data obtained for an electrolytic solution of ${\mbox{CdCl}}_2 {\mbox{H}}_2{\mbox{O}}$.
\end{abstract}
\maketitle

\section{Introduction}

Electrochemical processes have a broad and important range of applications, being the suitable context for introducing electrochemical impedance measurements techniques \cite{i1,i2,i3,i4,i5,i6,i7}. Indeed, the immittance or electrical impedance spectroscopy (IS) is a powerful method of characterizing many of the electrical properties of materials. It has been used to investigate the dynamics of bound or mobile charge in the bulk or interfacial regions of any kind of liquid or solid material (ionic, semiconducting, mixed electronic-ionic and even insulators (dielectrics)) \cite{i8}. The popularity of this tool in the research and development of materials is due to the fact that it involves a relatively simple electrical measurement whose results can be correlated with complex variables of materials such as, for example, mass transfer, rates of chemical reactions, dielectric properties, compositional influences on the conductance of solids, and others.

The analysis of the influence of the ions on the IS response of a given sample can usually be performed by solving the continuity equations, for the positive and negative ions, satisfying the Poisson's equation requirement for the electric potential across the sample (PNP model). These equations have to be solved with the boundary conditions on the electrodes and on the electric potential.  The first PNP analysis, involving completely blocking electrodes, was presented by Macdonald in 1953~\cite{ii9}. Later, an extension of this analysis involving a more complete theoretical model was presented~\cite{ii10}, followed by a general model proposed by Macdonald and Franceschetti~\cite{ii11}. In both extensions, the assumption of completely blocking electrodes has been modified by assuming different types of boundary conditions on the electrodes.
Anyway, the assumption that the electrodes are perfectly blocking and therefore the ionic current densities vanish on the electrodes can be successfully applied to the high-frequency region of the IS response~\cite{i9,i10}. However, it is far from being realistic in the low frequency region where different boundary conditions~\cite{i11}, namely,  reaction terms connected to generation and recombination of ions,  have been considered in the context of the usual PNP model to investigate the IS response in the linear approximation. Actually, several results obtained for IS response, for example, in nanostructured iridium oxide~\cite{i12}, water~\cite{i13,i14}, morphology and ion conductivity of gelatin -- LiClO$_{4}$  films~\cite{i15}, and ionic solutions~\cite{i16} have evidenced that the usual framework may not be suitable enough to describe the behavior of the system in the entire frequency range. Consequently, more complicated results in the low frequency domain have motivated the study of alternative models which extend the usual approach  to face these situations.

In electrolytic cells, the anomalous response that generalizes the Warburg model for the electrical impedance was proposed by Macdonald in 1985~\cite{i17}. Bisquert and co-workers~\cite{i18,i19,i20,i21} employing fractional calculus have investigated several situations with the purpose of determining the electrochemical impedance in this context. In Ref.~\cite{i22}, this approach was worked out by including the requirement to satisfy Poisson's equation. Another fractional -- type diffusional response for regions of finite length has been proposed in~\cite{i23}  (see also Ref.~\cite{ii23}) leading to an alternative model for the electrical impedance whose form is different from the ones treated in Ref.~\cite{i22}. In Ref.~\cite{i24},  the model investigated in~\cite{i22} for the electrical response of an electrolytic cell was revisited and extended by incorporating boundary conditions stated in terms of an integro-differential equation containing a temporal kernel. This mathematical way to state the boundary conditions permits one to perform a suitable choice for this temporal kernel in order to treat or to cover various different scenarios in a phenomenological perspective.

In this paper, the aim is to establish a general and unified mathematical framework to investigate formal aspects of the IS response obtained from a fractional diffusion equation of distributed order, with a reaction term, coupled to the Poisson's equation for the electrical potential governing the behavior of the ions in the bulk, and subjected to a general boundary condition. This boundary condition is built here in such a way to embody a large number of processes involving the surface (electrodes) and the bulk (sample) such as the charge transfer and the adsorption-desorption phenomenon at the electrode surfaces. After obtaining the exact solutions for the fundamental equations of this anomalous PNP (PNPA) model,  that fulfil the requeriments imposed by the boundary conditions,  we obtain an exact analytical expression for the impedance that contains some relevant expression already published as particular cases. In addition, however, this general expression points towards results never explored yet in the important framework of the PNP model. To explore part of these new findings, we promote a comparison between the theoretical predictions with a set of experimental data obtained by us for the electrolytic solution of ${\mbox{CdCl}}_2 {\mbox{H}}_2{\mbox{O}}$. The agreement is very good and illustrates the potential usefulness of this new formalism to analyze experimental impedance data expressed in the complex plane.

\section{Fundamental Equations: Fractional Diffusion and Reaction Term}

The bulk densities of ions $n_{\alpha}$  ($\alpha=+$ for positive and $\alpha =-$  for negative ones) are governed by the following fractional diffusion equation of distributed order:

\begin{widetext}
\begin{eqnarray}\label{e1}
{\cal{A}}\frac{\partial}{\partial t}n_{\alpha}(z,t)+{\cal{B}}\frac{\partial^{\gamma}}{\partial t^{\gamma}}n_{\alpha}(z,t)=-\frac{\partial}{\partial z}j_{\alpha}(z,t)
-\int_{-\infty}^{t}\zeta(t-t')n_{\alpha}(z,t')dt',
\end{eqnarray}
\end{widetext}
where $\gamma$  is the index of the fractional time derivative defined in the interval $0<\gamma<2$, in order to cover sub- and super-diffusive situations,  where ${\cal{A}}$  is dimensionless, while the dimension of ${\cal{B}}$   is given in terms of $t^{\gamma-1}$. In Eq.~(1), the reaction term can be connected to reaction diffusion processes and anomalous diffusion~\cite{i25,i26,i27}. Similar terms have also been used to build neural models of pattern formation where nonlocal effects play an important role~\cite{i28,i29}. Typical situations connected to the previous equation may present different diffusive regimes such as the ones discussed in Refs.~\cite{i31,i30} or the ones presented in
Refs.~\cite{i32,i33,i34,i35}, characterized by a finite phase velocity. The fractional operator considered here is the Caputo's one, given by:

\begin{eqnarray}
\frac{\partial^{\gamma}}{\partial t^{\gamma}}n_{\alpha}(z,t)=\frac{1}{\Gamma\left(k-\gamma\right)}\int_{t_{0}}^{t}dt'\frac{n_{\alpha}^{(k)}(z,t)}{(t-t')^{\gamma-k+1}},
\end{eqnarray}
with $k-1<\gamma<k$  and $n_{\alpha}^{(k)}(z,t)\equiv
\partial^{k}_{t}n_{\alpha}(z,t)$. As discussed in Ref.~[36], it is useful to consider $t_{0}\rightarrow -\infty$  in order to analyze the response of the system to a periodic applied potential like the one to be introduce in the next Section.  The drift-diffusion current density is given by:
\begin{eqnarray}
\label{current} j_{\alpha}(z,t)= -{\cal{D}}_{\alpha}\frac{\partial }{\partial z}n_{\alpha}(z,t)\mp \frac{q {\cal{D}}_{\alpha}}{k_{B}T}n_{\alpha}(z,t)\frac{\partial }{\partial z}V(z,t),
\end{eqnarray}
where  ${\cal{D}}_{\alpha}$  is the diffusion coefficient for the mobile ions (assumed hereafter as being the same for positive and negative ones) of charge $q$,   $V$ is the actual electric potential across a sample of thickness  $d$, with the electrodes placed at the positions $z=\pm d/2$  of a Cartesian reference frame in which $z$ is the axis normal to them, $k_B$  is the Boltzmann constant, and $T$, the absolute temperature. The effective time-dependent potential across the sample is determined by the Poisson's equation
\begin{equation}
\label{e3} \frac{\partial^{2}}{\partial z^{2}}V(z,t)=-\frac{q}{\varepsilon}\left[n_{+}(z,t)-n_{-}(z,t)\right]\;,
\end{equation}
where the dielectric coefficient is  $\varepsilon$ (measured in $\varepsilon_{0}$  units).

The description of the influence of the surface on the ions is strongly connected to the choice of the boundary condition to be satisfied by the solutions of Eq.~(1). Various electrode reaction rate boundary conditions suitable for mean-field PNP models have been recently investigated by Macdonald~\cite{i37} in order to generalize previous models from full to partial blocking of mobile charges at the electrodes. Here, to face a general situation, it is proposed to consider the solutions of Eq.~(1) subjected to the following boundary condition:

\begin{widetext}
\begin{eqnarray}
\label{e2} \left. j_{\alpha}(z,t)\right|_{z=\pm\,\frac{d}{2}}\!=\! \left. \pm k_{\alpha,e} E\left(z,t\right)\right|_{z=\pm\frac{d}{2}}\left.\pm \int_{0}^{1}\!\!\!d\overline{\vartheta}\;\widetilde{\tau}(\overline{\vartheta})\!\!
\int_{-\infty}^{t}\!\!\!\!\!\!\!d\overline{t} {\cal{K}}_{a}(t-\overline{t},\overline{\vartheta})\frac{\partial^{\overline{\vartheta}}}{\partial
\overline{t}^{\overline{\vartheta}}}n_{\alpha}\!\left(z,\overline{t}\right)\right|_{z=\pm\,\frac{d}{2}},
\end{eqnarray}
\end{widetext}
where $k_{\alpha,e}$ represent the parameters of the Ohmic model, measured, e.g. in $1/(V m s)$ in the SI system~\cite{i11}. Thus, the first term simply states that the ionic current density on the electrode is proportional to the surface electric field. The second term contains a temporal kernel ${\cal{K}}_{a}(t,\overline{\vartheta})$  convoluted with a fractional time derivative of the bulk density of ions, calculated on the electrodes.

This integro-differential expression for the boundary conditions has,  as particular cases,  many other physical situations considered in different approaches [11, 16, 24, 38] (completely or partially blocking electrodes, Ohmic and transparent electrodes, adsorption-desorption processes,  and many others). Built in this general form, Eq. (5) not only groups different known cases but also permits one to interpolate several contexts eventually playing relevant roles in the description of the electrical response of a given system. Moreover, for what concerns the second term of the boundary condition, it may formally be obtained, as well as the Eq. (1), in the context of the continuous time random walk \cite{i39,i40} if reactive boundary conditions were considered, similarly to the developments performed in~\cite{i41,i42}. In this sense, a particular choice for the kernel, as done in
Refs.~[16, 14, 43], may be dictated by the physical process manifested by the system to be considered. In this way as well, the set of equations (1)-(4), together with the boundary conditions (5), constitute a new mathematical statement for the usual PNP and anomalous PNPA diffusional models,  embodying a quite large class of particular situations considered before,  but also pointing out towards applications to novel scenarios in the field of electrochemical impedance.

\section{Analytical solutions for the small AC signal}

An analytical solution for the previous equations and, consequently, an expression for the electrical impedance embodying all the situations mentioned before can be found in the linear approximation (small a.c. signal limit). In this approximation, one may consider that  $n_{\alpha}(z,t)={\cal{N}}+\delta n_{\alpha}(z,t)$, with  ${\cal{N}}\gg \delta |n_{\alpha}(z,t)|$, where ${\cal{N}}$  represents the number of ions per unit volume in the bulk. For simplicity's sake, one chooses $\delta n_{\alpha}(z,t)=\eta_{\alpha}(z)e^{i\omega t}$ and $V(z,t)=\phi(z)e^{i\omega t}$  to analyze the impedance when the electrolytic cell is subjected to a time dependent potential in the form $V(\pm d/2,t) = \pm V_0 e^{i \omega t}/2$, where $V_0$ is the amplitude  and $\omega = 2\pi f$ is the frequency of the applied potential,  since the stationary state is reached. Substitution of these expressions into Eqs.~(1),~(3),~(4), and~(5) yields a set of coupled equations which may be simplified further by using the auxiliary functions $\psi_{-}(z)=\eta_{+}(z)-\eta_{-}(z)$  and $\psi_{+}(z)=\eta_{+}(z)+\eta_{-}(z)$. The first two equations become

\begin{eqnarray}
\label{e5}
\frac{d^{\,2}}{d z^{2}}\psi_{\pm}(z)&=&\nu_{\pm}^{2}\psi_{\pm}(z),
\end{eqnarray}
where   $\nu_{-}^{2}={\Lambda}\left(i
\omega\right)/{\cal{D}}+1/\lambda^{2}$
 and  $\nu_{+}^{2}={\Lambda}\left(i \omega\right)/{\cal{D}}$, in which ${\Lambda}\left(i
\omega\right)={\cal{A}}(i \omega)+ {\cal{B}}(i \omega)^{\gamma}+\zeta(i\omega)$, where  $\zeta(i\omega)=e^{-i\omega t}\int_{-\infty}^{t}dt'\zeta(t-t')e^{i\omega t'}$  and  $\lambda = \sqrt{\varepsilon k_B T/(2 {N}q^2)} $ is the Debye's screening length. The other two equations connected to the boundary conditions become

\begin{widetext}
\begin{eqnarray}
\label{e6}
\left. \left({\cal{D}}\frac{d }{d z}\psi_{-}(z)+ \frac{2q {\cal{N}}}{k_{B}T}{\cal{D}}
\frac{d }{d z}\phi(z)\right)\right|_{z=\pm \frac{d}{2}}&=&\left.\pm \overline{k}_{e}\frac{d}{dx}\phi(z)\right|_{z=\pm\frac{d}{2}} \mp \left.\Upsilon(i\omega) \psi_{-}(z)\right|_{z=\pm \frac{d}{2}}, \\
\label{e61}
\left.{\cal{D}}\frac{d }{d z}\psi_{+}(z)\right|_{z=\pm \frac{d}{2}}&=&\mp \left.\Upsilon(i\omega)\psi_{+}(z)\right|_{z=\pm \frac{d}{2}},
\end{eqnarray}
\end{widetext}
with  $\Upsilon_{a}(i\omega)=e^{-i\omega t}\int_{0}^{1}d\vartheta\widetilde{\tau}(\vartheta)\left(i\omega\right)^{\vartheta}
\int_{-\infty}^{t}d\overline{t}\,{\cal{K}}_{a}(t-\overline{t})e^{i\omega \overline{t}}$ and  $ \overline{k}_{e}= k_{+,e}-k_{-,e}$. The solutions of Eqs. (6) are  $\psi_{\pm}(z)={\cal{C}}_{\pm,1}e^{\nu_{\pm}z}+{\cal{C}}_{\pm,2}e^{-\nu_{\pm}z}$, where  ${\cal{C}}_{\pm,1}$ and ${\cal{C}}_{\pm,2}$ have to be determined by the boundary conditions and by invoking the symmetry of the potential, i.e., $V(z,t)=-V(-z,t)$, which implies ${\cal{C}}_{-,1}=-{\cal{C}}_{-,2}$ and, consequently,
\begin{eqnarray}
\label{e11}
\psi_{-}(z)&=&2{\cal{C}}_{-,1}\sinh\left(\nu_{-}z\right),  \\
\phi(z)&=&-\frac{2q}{\varepsilon\nu_{-}^{2}} {\cal{C}}_{-,1}\sinh\left(\nu_{-}z\right)+\overline{{\cal{C}}}z,
\end{eqnarray}
with $\overline{{\cal{C}}}$  determined by means of the boundary conditions, similarly to  ${\cal{C}}_{\pm,1}$ and  ${\cal{C}}_{\pm,2}$. After some calculation, it is possible to show that the impedance, for the case discussed here, is given by the general analytical expression
\begin{eqnarray}
\label{impedance}
{\cal{Z}}=\frac{2}{{\cal{S}}\varepsilon\nu_{-}\Delta(i\omega)} \left[\frac{1}{\lambda^2\nu_{-}}\tanh\left(\nu_{-}d/2\right)+\frac{d}{2{\cal{D}}}{\cal{E}}(i\omega)\right],
\end{eqnarray}
in which
\begin{eqnarray}
\Delta(i\omega)&=&\left(1/\lambda^2+\chi\left(\Lambda(i\omega)+\omega_{e}\right)/{\cal{D}}\right)\left(i\omega+\omega_{e}\right)/\nu_{-}\nonumber \\ &+&
\left(1/\lambda^2+\chi\left(i\omega+\omega_{e}\right)/{\cal{D}}\right)\Upsilon(i\omega)
\tanh\left(\nu_{-}d/2\right), \nonumber
\end{eqnarray}
with  ${\cal{E}}(i\omega)=\chi\left[\Lambda(i\omega)+\omega_{e}\right]+\chi\nu_{-}\Upsilon(i\omega)\tanh
\left(\nu_{-}d/2\right)$,  $\chi=1/\left[1-\lambda^{2}\overline{k}_{e}q/\left({\cal{D}}\varepsilon\right)\right]$,   $\omega_{e}=\overline{k}_{e}q/\varepsilon$, and  ${\cal{S}}$ is the electrode area. It is worth mentioning that Eq. (11) extends the results found in Refs. \cite{i38,i22,i16,i24} to an even more general context, with the surface effects described by the boundary condition given by Eq. (5). Indeed, all the cases discussed in these works can now be considered as particular cases of the present approach.

\begin{figure}
 \centering \DeclareGraphicsRule{ps}{eps}{}{*}
\includegraphics*[scale=.35,angle=0]{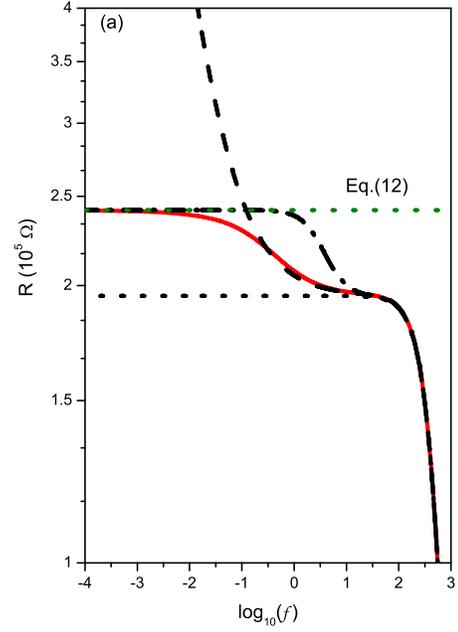}
\includegraphics*[scale=.35,angle=0]{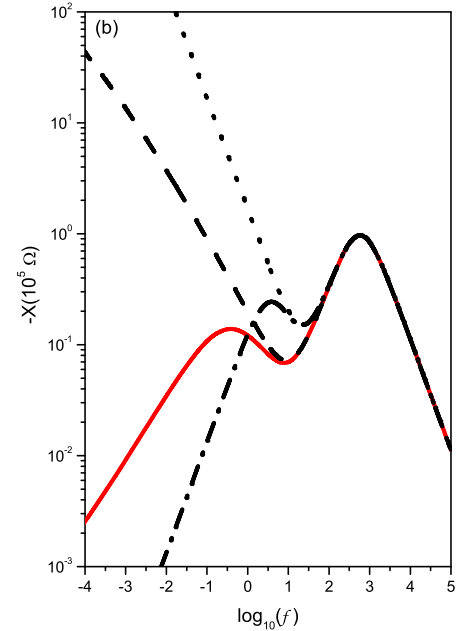}
\caption{
(a) The behavior of the real,  ${\mbox{R}}={\mbox{Re}}\,{\cal{Z}}$, and (b)  the imaginary,  ${\mbox{X}}={\mbox{Im}}\,{\cal{Z}}$, parts of the impedance versus the frequency in absence of the reaction term. The red solid line corresponds to the case $\kappa_{a,1}\neq\kappa_{a,2}\neq 0$ and $\overline{k}_{e}\neq 0$. The black dashed line is the case $\kappa_{a,1}\neq\kappa_{a,2}\neq 0$ with $\overline{k}_{e}= 0$. The black dotted line is the case $\kappa_{a,1}=\kappa_{a,2}= 0$ and $\overline{k}_{e}= 0$. The black dashed-dotted line is characterized by $\kappa_{a,1}=\kappa_{a,2}= 0$ and $\overline{k}_{e}\neq 0$. For simplicity, we consider the following values for the parameters:  ${\cal{S}}=2\times10^{-3}\;m^{2}$, $\varepsilon= 80\varepsilon_{0}$ ($\varepsilon_{0}= 8.85\times10^{-12}\; C^{2}/(Nm^{2})$), $\gamma=1$, ${\cal{D}}=2\times 10^{-9}\;m^{2}/s$, $d=10^{-3}m$, $\kappa_{a,1}= 10^{-5}\; m/s$, $\kappa_{a,2}= 1.95\times10^{-6}\; m/s$, $\overline{k}_{e}= 10^{11} (msV)^{-1}$, $\lambda=7.437\times 10^{-7}m$,  $\tau=0.01 s$, $\eta_{1}= 0.6$, and $\eta_{2}=0.31$. } \label{Figura1}
\end{figure}

\section{Results for charge transfer and adsorption processes}

Figures~\ref{Figura1}a  and~\ref{Figura1}b show the behavior of the real and imaginary parts of Eq. (11) in absence of the reaction term, but considering the presence of charge transfer, i.e.,  $\overline{k}_{e}\neq 0$, and an adsorption process characterized by  $\Upsilon(i\omega)=\kappa_{a,1}\tau(i\omega\tau)^{1-\eta_{1}}+\kappa_{a,2}\tau(i\omega\tau)^{1-\eta_{2}}$. Note that the choice for $\Upsilon(i\omega)$  essentially describes the presence of a double adsorbing layer, each one characterized by an effective thickness ($\kappa_{a,1}\tau$ and $\kappa_{a,2}\tau$) and a frequency dependence (represented by the exponents $\eta_{1}$ and $\eta_{2}$) to account for the different interactions along these layers. The dashed line is obtained for  $\overline{k}_{e}= 0$. The dotted line is the standard model of blocking electrodes, i.e.,  $\overline{k}_{e}= 0$ and  $\kappa_{a,1}=\kappa_{a,2}=0$. The red line combines the situation discussed for ${\cal{K}}_{a}(i\omega)$  with  $\overline{k}_{e}\neq 0$  and the black dashed-dotted line is the case  $\kappa_{a,1}=\kappa_{a,2}=0$ with $\overline{k}_{e}\neq 0$. The results of Fig.~\ref{Figura1}a show that in the low frequency domain the effect of charge transfer governs the electrical response of the system,  even in the presence of an adsorption-desorption process. This feature is also manifested in the imaginary part of the impedance,  as illustrated in Fig.~\ref{Figura1}b. As a consequence, one can notice the presence of a second plateau in the real part of the impedance,  in the low frequency domain,  while the imaginary part of the impedance decreases and goes to zero in this limit. Thus, the influence of the processes governed by the previous  $\Upsilon(i\omega)$, when the charge transfer is present, i.e.,  $\overline{k}_{e}\neq 0$, is absent in the low frequency domain. In fact, this result can be verified for the adsorption-desorption processes characterized by  $\Upsilon(i\omega)=\kappa_{a}\tau(i\omega\tau)^{1-\delta}$, with $\omega\rightarrow 0$  and $0<\delta<1$. This point is evidenced by the results illustrated in Fig.~\ref{Figura1} and by the asymptotic (analytical) results obtained from Eq.~(11). Specifically, for the real part of the impedance, one has

\begin{eqnarray}
{\mbox {Re}}\,{\cal{Z}}\sim \frac{2\lambda}{\overline{k}_{e}q{\cal{S}}}\left[1+\frac{\overline{k}_{e}q}{{\cal{D}}\varepsilon}\lambda^{2}\left(\frac{d}{2\lambda}-1\right)\right]
\end{eqnarray}
which only involves the constants connected to the bulk and to the charge transfer.

\begin{figure}
\centering \DeclareGraphicsRule{ps}{eps}{}{*}
\includegraphics*[scale=.35,angle=0]{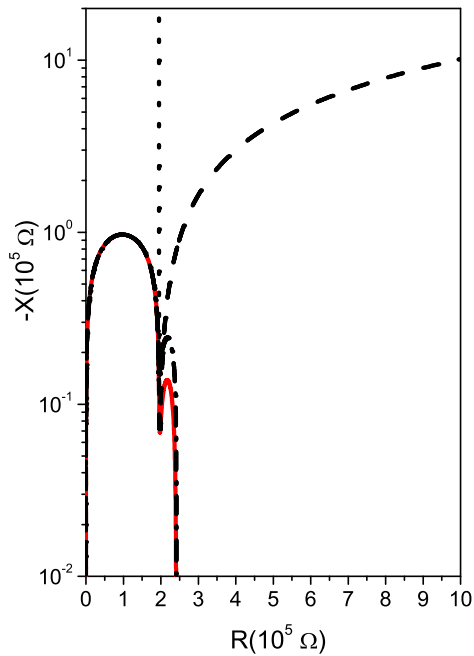}
\caption{Behavior of the $-{\mbox{X}}$ versus ${\mbox{R}}$ for the cases discussed in Fig. (\ref{Figura1}) in order to evidence
the differences in the low frequency limit.} \label{Figura2}
\end{figure}

In Fig.~\ref{Figura2}, the Nyquist plot of the cases shown in Fig.~\ref{Figura1} is presented. Notice that the main difference, as discussed in Fig.~\ref{Figura1}, is found in the low frequency limit. The first semicircle can be connected to the bulk effects and the second part of Fig.~\ref{Figura2} can be related to the effects present in the low frequency limit, i.e., the surface effects. In this sense, the cases characterized by  $\overline{k}_{e}\neq 0$  present an additional semicircle which is not present when ${\cal{K}}_{a}(i\omega)\sim 1/(i\omega\tau)^{\delta}$  or ${\cal{K}}_{a}(i\omega)=0$,  with  $\overline{k}_{e}=0$. These considerations may be useful when one investigates experimental data and can be used to establish a relation between Eq.~(11) and an approach involving equivalent circuits.
	
	         At this point, it is mandatory to underline that a lot of physical situations (see, for example, Refs.~\cite{i13,i14,i15,i16}) present, in the low frequency limit, an asymptotic behavior governed by  ${\cal{Z}}\sim 1/(i\omega)^{\sigma}$,  with  $0<\sigma<1$,  which is the behavior essentially exhibited by Eq.~(11),  for  $\omega\rightarrow 0$,  when ${\cal{K}}_{a}(i\omega)\sim 1/(i\omega\tau)^{\delta}$  for  $\overline{k}_{e}=0$. This behavior is absent if the standard model, i.e., usual PNP diffusional model,  is considered,  and has been also connected to the roughness of the surface by some authors~\cite{i44,i45} (see also Refs.~\cite{i46,i47}, for additional discussions).

	The adsorption process may have a pronounced influence when  $\overline{k}_{e}\neq 0$, in the low frequency limit,  if the second term of Eq. (5) has the following asymptotic behavior: $\Upsilon_{a}(i\omega)$ $\sim$ $\kappa\tau$. Actually, a typical situation can be found if one uses the generalized Chang-Jaffé boundary conditions accounting for specific ion adsorption at the interfaces~\cite{i10}. The original Chang-Jaffé electrode-reaction boundary conditions~\cite{i48} were introduced by Friauf~\cite{i49} to investigate partial-blocking effects. Subsequently, in Ref.~\cite{i50}, extended Chang-Jaffé boundary conditions have been considered and were also generalized to include specific ion adsorption,  a few years later~\cite{ii11} (for a detailed discussion see Ref.~\cite{i37}). These boundary conditions can be implemented in the present framework by incorporating $\kappa_{CJ}\tau$  into  $\Upsilon_{a}(i\omega)$, where $\kappa_{CJ}$  is a single Chang-Jaffé parameter. This scenario leads one to obtain

\begin{eqnarray}
{\mbox {Re}}\,{\cal{Z}}\sim \frac{\left(2{\cal{D}}+d\kappa_{CJ}\tau\right)\varepsilon\lambda^{2}+2\overline{k}_{e}q\lambda^{3}\left[d/\left(2\lambda\right)-1\right]}
{{\cal{S}}\overline{k}_{e}{\cal{D}}q\varepsilon\left[1+\varepsilon\kappa_{CJ}\tau/\left(\overline{k}_{e}q\lambda\right)\right]}\;.
\end{eqnarray}

\begin{figure}
\centering \DeclareGraphicsRule{ps}{eps}{}{*}
\includegraphics*[scale=.35,angle=0]{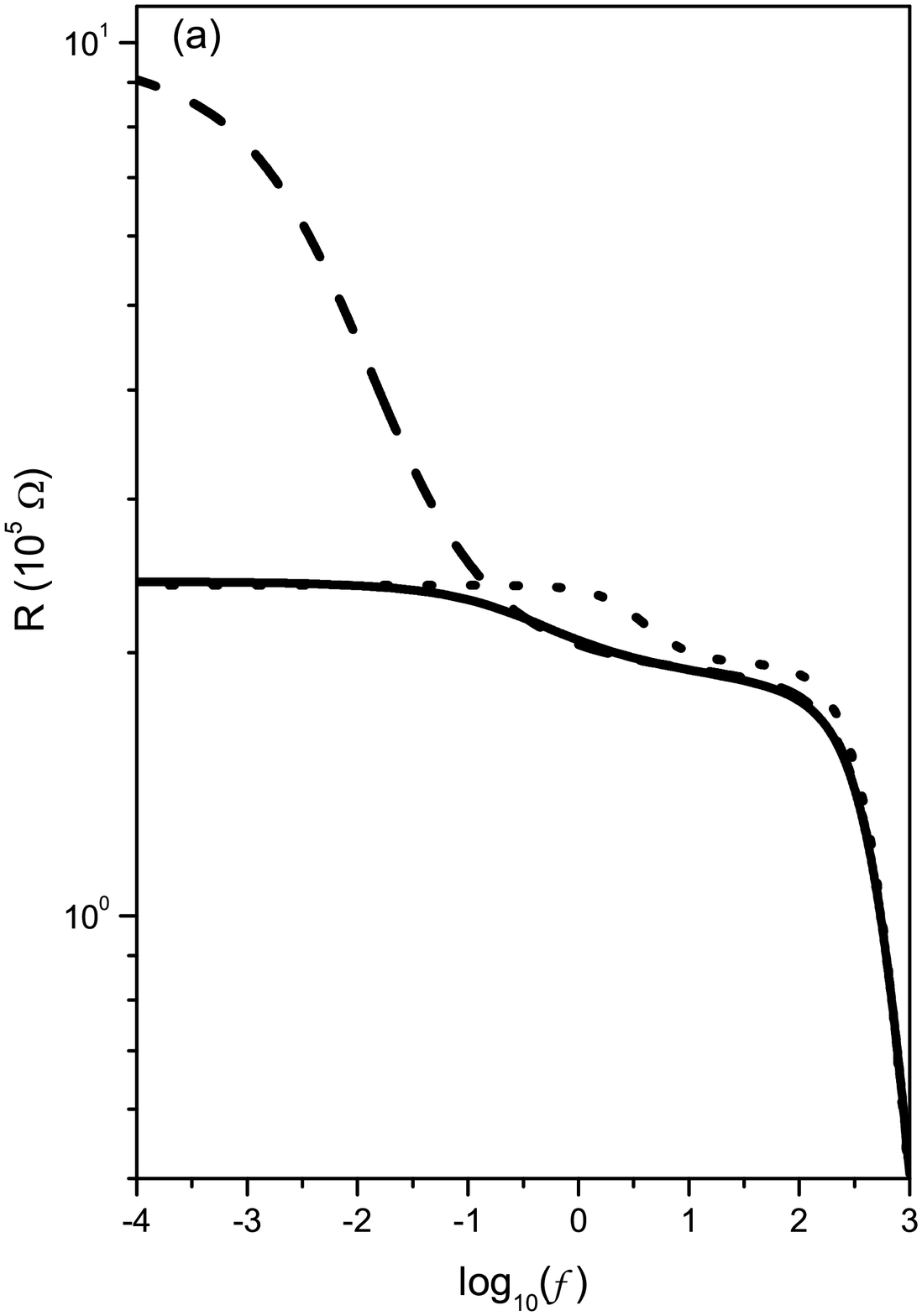}
\includegraphics*[scale=.35,angle=0]{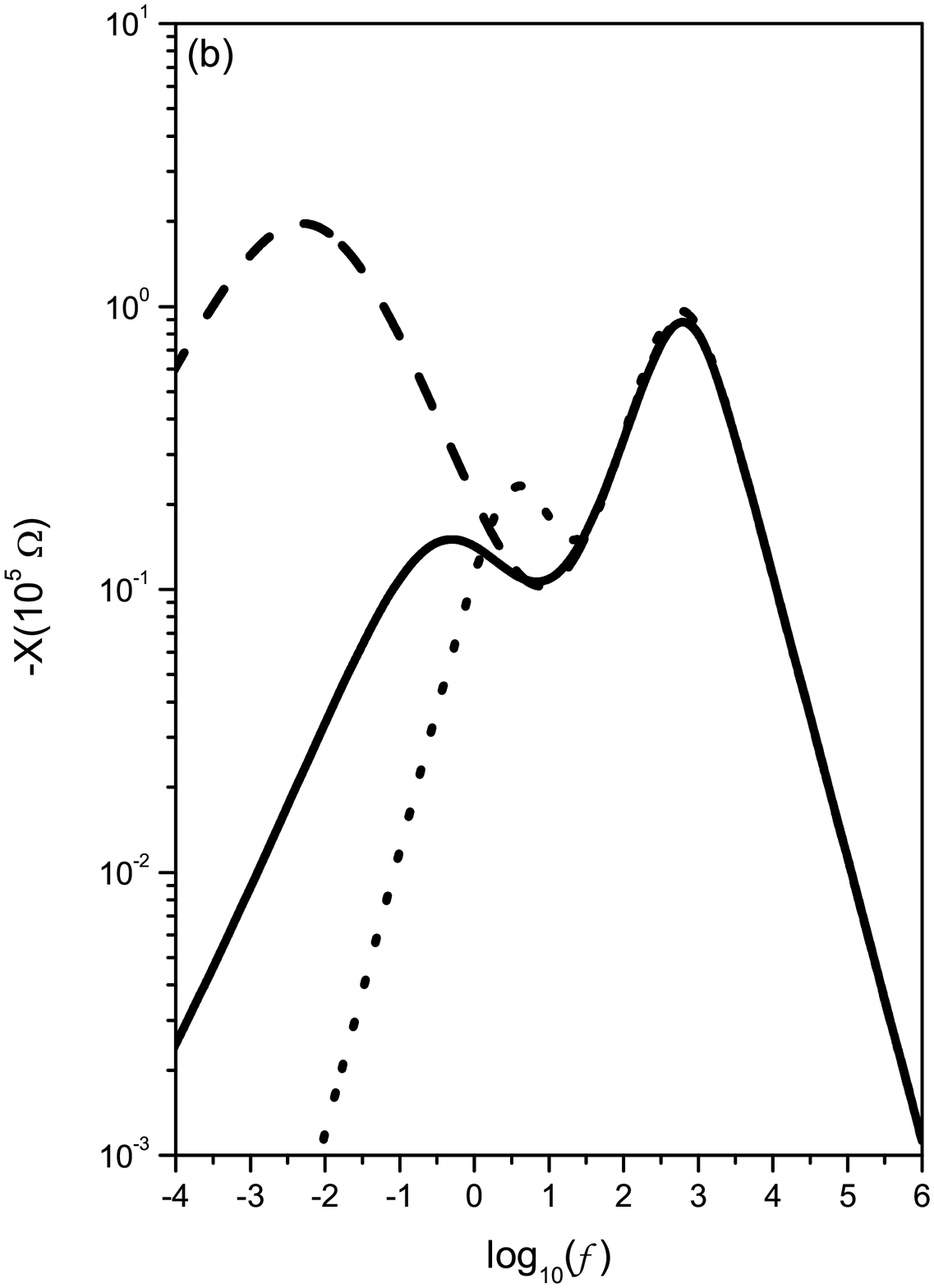}
\caption{(a) The behavior of the real,  ${\mbox{R}}={\mbox{Re}}\, {\cal{Z}}$, and (b)  the imaginary,  ${\mbox{X}}={\mbox{Im}}\, {\cal{Z}}$, parts of the impedance versus the frequency in absence of the reaction term, respectively. The black solid line corresponds to the case $\gamma=0.7$, $\kappa_{a,1}\neq\kappa_{a,2}\neq 0$, $\kappa_{CJ}=0$, and $\overline{k}_{e}\neq 0$. The black dashed line is the case $\gamma=0.8$, $\kappa_{a,1}\neq\kappa_{a,2}\neq 0$, $\kappa_{CJ}\neq0$ with $k_{e}= 0$. The black dotted line is the case $\gamma=0.9$, $\kappa_{a,1}=\kappa_{a,2}= 0$, $\kappa_{CJ}\neq0$, and $\overline{k}_{e}\neq 0$. The black dashed-dotted line is characterized by $\kappa_{a,1}=\kappa_{a,2}= 0$, $\kappa_{CJ}=0$ and $k_{e}\neq 0$. The figures were drawn for the following values of the parameters:  ${\cal{S}}=2\times10^{-3} \;m^{2}$, $\varepsilon= 80\varepsilon_{0}$, ${\cal{D}}=2\times 10^{-9}\; m^{2}/s$, $d=10^{-3}\;m$, $\kappa_{a,1}= 10^{-5} \; m/s$, $\kappa_{a,2}= 1.95\times10^{-6}\;m/s$, $\overline{k}_{e}= 10^{11}\;(msV)^{-1}$, $\lambda=7.43 \times 10^{-8}\;m$, $\kappa_{CJ}= 10^{-4} \; m/s$,  $\tau=0.01\;s$, ${\cal{A}}=0.8$, $\eta_{1}= 0.6$, and $\eta_{2}=0.31$. } \label{Figura3}
\end{figure}

\begin{figure}
\centering \DeclareGraphicsRule{ps}{eps}{}{*}
\includegraphics*[scale=.35,angle=0]{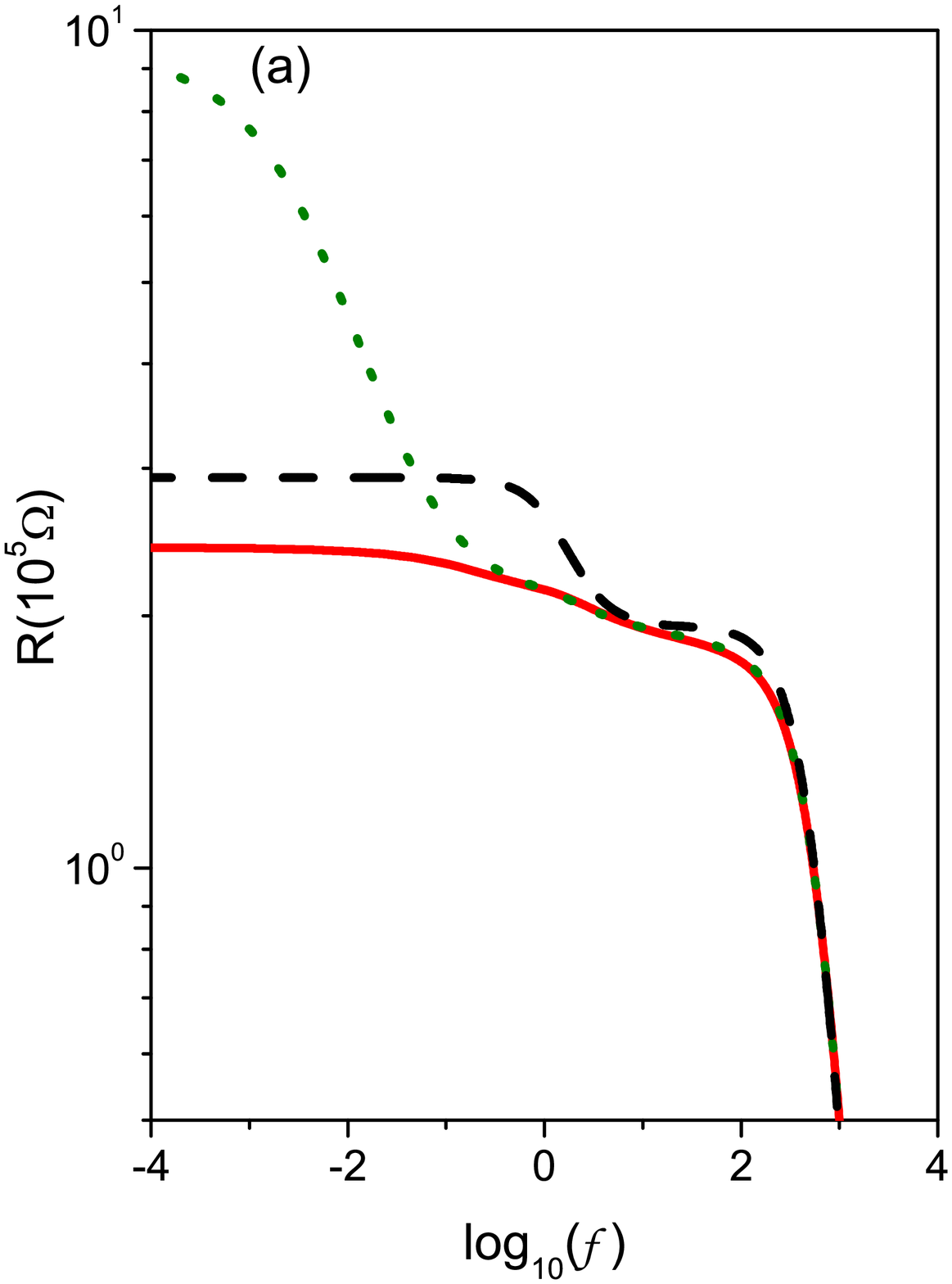}
\includegraphics*[scale=.35,angle=0]{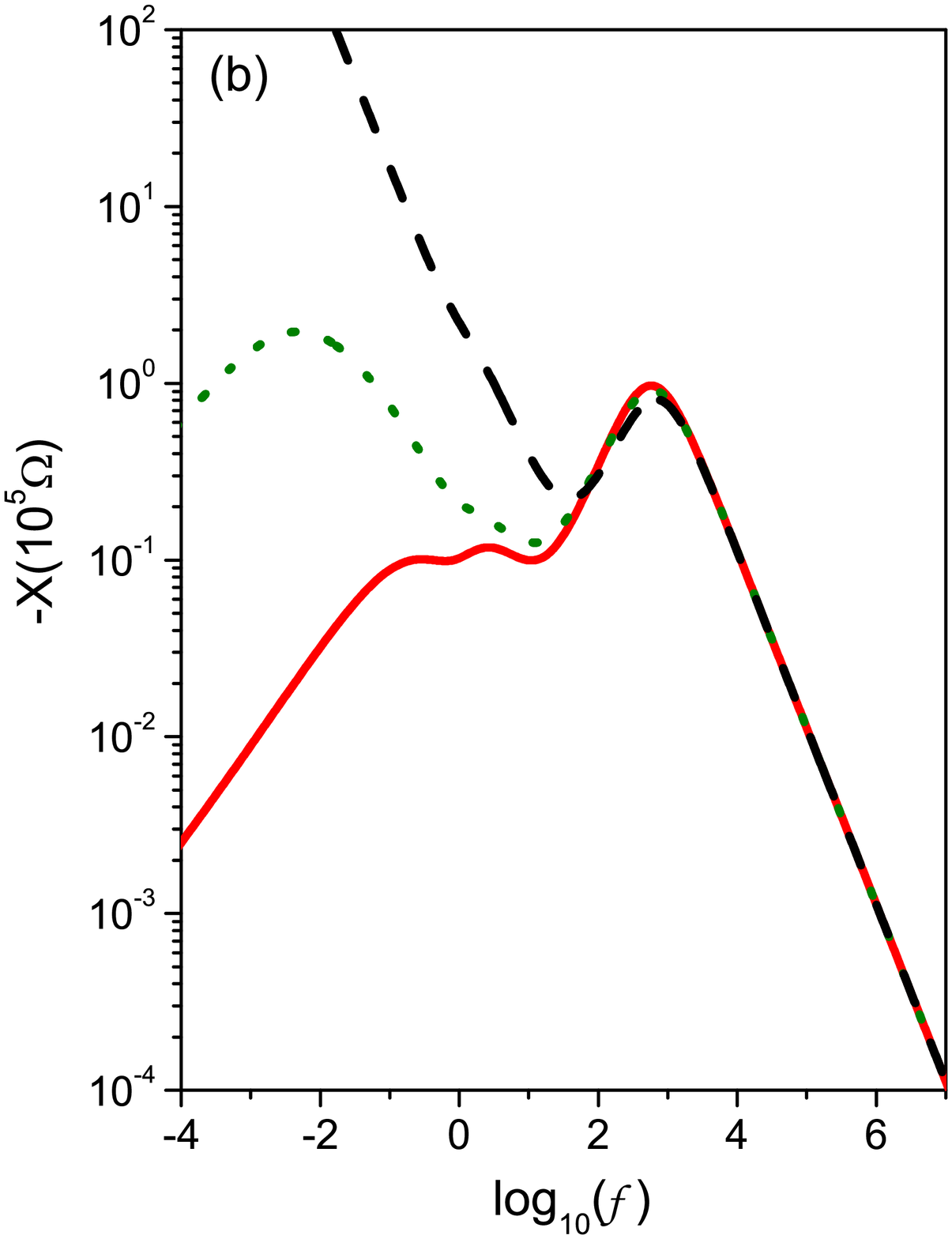}
\caption{The behavior of the real,  ${\mbox{R}}={\mbox{Re}}\, {\cal{Z}}$, (a) and the imaginary,  ${\mbox{X}}={\mbox{Im}}{\cal{Z}}$ (b) parts of the impedance versus the frequency with the reaction term, chosen to be given by $\zeta(i\omega) = 5 i\omega/(10 + i\omega)$, which can formally be connected to process of generation and recombination of ions [25, 27]. The red dashed line corresponds to the case $\gamma=0.7$, $\kappa_{a,1}\neq\kappa_{a,2}\neq 0$, $\kappa_{CJ}=0$, and $\overline{k}_{e}\neq 0$. The green dotted line is the case $\gamma=0.8$, $\kappa_{a,1}\neq\kappa_{a,2}\neq 0$, $\kappa_{CJ}\neq0$ with $k_{e}= 0$. The black solide line is the case $\gamma=1$, $\kappa_{a,1}=\kappa_{a,2}\neq 0$, $\kappa_{CJ}\neq0$, and $\overline{k}_{e}= 0$. The black dashed-dotted line is characterized by $\kappa_{a,1}=\kappa_{a,2}= 0$, $\kappa_{CJ}=0$ and $k_{e}\neq 0$. For simplicity, we consider the following values for the parameters:  ${\cal{S}}=2\times10^{-3} \;m^{2}$, $\varepsilon= 80\varepsilon_{0}$, ${\cal{D}}=2\times 10^{-9}\; m^{2}/s$, $d=10^{-3}\;m$, $\kappa_{a,1}= 10^{-5} \; m/s$, $\kappa_{a,2}= 1.95\times10^{-6}\;m/s$, $\overline{k}_{e}= 10^{11}\;(msV)^{-1}$, $\lambda=7.43 \times 10^{-8}\;m$, $\kappa_{CJ}= 10^{-4} \; m/s$,  $\tau=0.01\;s$, ${\cal{A}}=0.8$, $\eta_{1}= 0.6$, and $\eta_{2}=0.31$. } \label{Figura4}
\end{figure}

Figures~\ref{Figura3}a and~\ref{Figura3}b illustrate the behavior of the impedance for   $\Upsilon_{a}(i\omega)$ $=$ $\kappa_{CJ}\tau$ $+$ $\overline{\Upsilon}_{a}(i\omega)$,        where    $\overline{\Upsilon}_{a}(i\omega)$ $=$ $\kappa_{a,1}\tau_{1}(i\omega\tau_{1})^{1-\eta_{1}}$ $+$ $\kappa_{a,2}\tau_{2}(i\omega\tau_{2})^{1-\eta_{2}}$, with $\zeta(i\omega)=0$  and  $\overline{k}_{e}\neq 0$. Notice that, in the low frequency limit, the behavior of the real and imaginary parts of the impedance is governed by the Chang-Jaffé condition and the charge transfer.  Figure~\ref{Figura4} illustrates the effects produced on the electrical response by the reaction term. In this case, the reaction term has influence on the behavior of the electrical response in the low frequency limit.

\begin{figure}
 \centering \DeclareGraphicsRule{ps}{eps}{}{*}
\includegraphics*[scale=.35,angle=0]{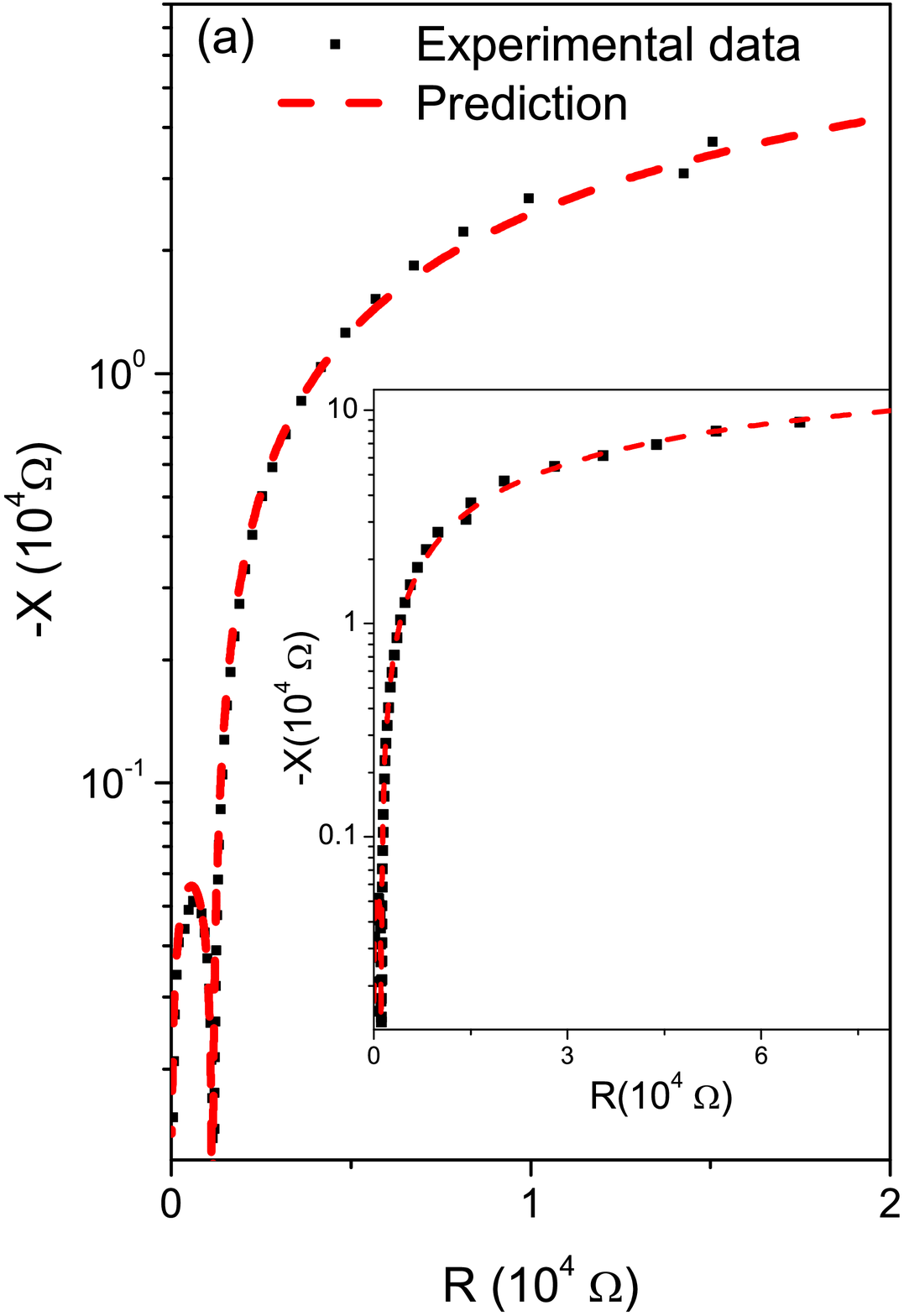}
\includegraphics*[scale=.35,angle=0]{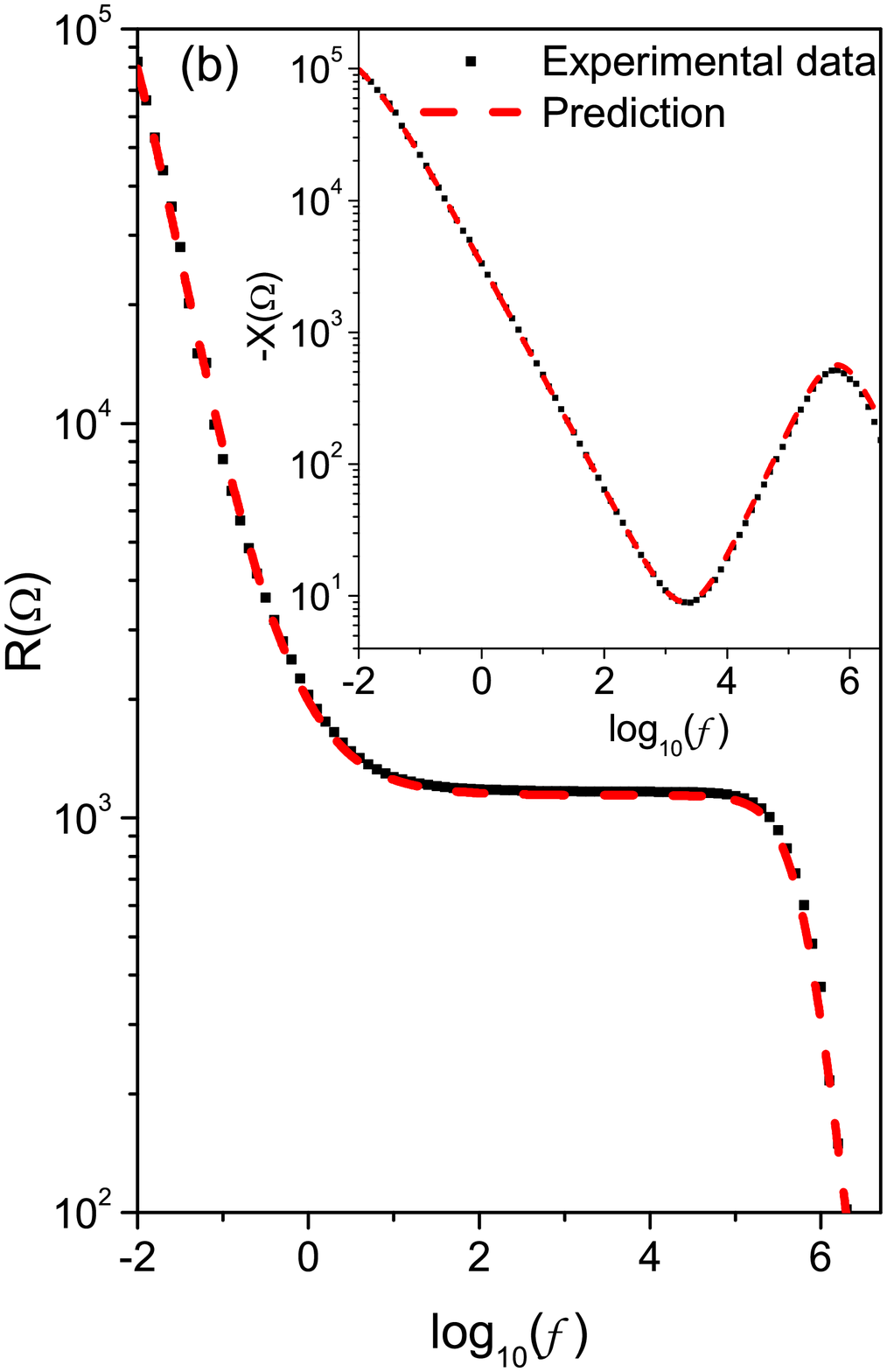}
\caption{
Comparison of the experimental data with the predictions of the model presented here for the real, ${\mbox{R}}={\mbox{Re}}{\cal{Z}}$ (a), and imaginary, ${\mbox{X}}={\mbox{Im}}{\cal{Z}}$ (b), parts of the impedance. A good agreement between the experimental data and the predictions was obtained for the following values of the parameters:   $S=3.14 \times 10^{-4}\;m^2$, $\epsilon=80.03\epsilon_{0}$ ($\varepsilon_{0}= 8.85\times10^{-12}\; C^{2}/(Nm^{2})$), $D=3.05 \times 10^{-9}\; m^2/s$, $d=10^{-3}m$, $\gamma=0.95$, ${\cal{A}}=0.98$, $\kappa_{a,1}=8.67 \times 10^{-5} \;m/s$, $\kappa_{a,2}=6.24\times 10^{-7} m/s$, $\lambda=2.80\times 10^{-8} m$, $\tau=1.64\times 10^{-3}s$, $\eta_{1}=0.158$, and $\eta_{2}=0.899$.
Note that the inset in Fig. 5a shows the agreement between the phenomenological model proposed here and the experimental data in the low frequency limit in according to Fig. 5b.} \label{Figura5}
\end{figure}

Finally, in order to illustrate how the formalism presented here works when applied to a specific experimental context, the theoretical predictions for ${\cal{Z}}$ will be compared with the experimental data obtained for an electrolytic solution of ${\mbox {CdCl}}_{2}{\mbox{H}}_{2}{\mbox{O}}$  (supplied by QEEL - Indústrias Qumicas S.A. with over 99.9 as received) dissolved in Milliq water, concentration  $8.75\times10^{-3}{\mbox{mol}}\;{\mbox{L}}^{-1}$.  Following the procedure described in Refs. [14, 16], the measurements of real and imaginary parts of the impedance were performed by using a Solartron SI 1296 A impedance/gain phase analyzer. The frequency range used was from  $10^{-2}$ to $10kHz$ . The amplitude of the AC applied voltage was $20 mV$ and temperature is $298K$. The ionic solutions were placed between two circular surfaces spaced $1.0 mm$ from each other. The area of electrical electrodes was $3.14 cm^2$. We used the electrical contact of stainless steel. Before starting the measurements, we adopted the following cleaning procedure: first, the electrodes were washed with detergent and deionized water also polished with fine sandpaper. Then, the electrodes were placed on ultrasonic bath for 10 min. After performing this procedure, the ionic solution, formed by the  Milli-Q water with a quantity of the salt ${\mbox {CdCl}}_{2}{\mbox{H}}_{2}{\mbox{O}}$  completely dissolved, is introduced in the electrodes ($1.0 mm$ thickness) and the real and the imaginary part of the impedance are measured. Both, the experimental data and the theoretical predictions, are presented in
Figs.~\ref{Figura5}a and~\ref{Figura5}b. Note that a suitable choice of the boundary conditions,  to describe the experimental data,  can be performed by analyzing the Nyquist plot of the experimental data and comparing it with the predictions show in
Fig.~\ref{Figura2}. The behavior of imaginary part of the impedance also plays an important role on the choice of the boundary condition. In fact, as it
was shown in Ref.~\cite{i52}, in the limit of low frequency limit, it can be directly connected to the surface effects and, consequently, with the choice of the boundary condition. These comparisons lead one to the choice   $\Upsilon(i\omega)$ $=$ $\kappa_{a,1}\tau(i\omega\tau)^{1-\eta_{1}}$ $+$ $\kappa_{a,2}\tau(i\omega\tau)^{1-\eta_{2}}$ with  $\overline{k}_{e}=0$.  This particular choice accounts,  in the range of frequency considered in Fig.~\ref{Figura5}, for an adsorption-desorption-like process occurring at the electrodes limiting the electrolytic solution, with $\kappa_{a,1}\tau$  and $\kappa_{a,2}\tau$  being two effective thicknesses.  They account for the spatial extensions of two different layers near to the electrode surface,  in which the effective interaction changes behavior. More precisely, the behavior of the interaction in each of these layers is governed, in the frequency domain,  by the exponents  $\eta_{1}$  and  $\eta_{2}$,  which, in turn, tell us how these layers interplay to build an effective diffusive layer in the neighborhood of the electrodes, in a phenomenological  perspective. Finally, it also should be mentioned  that to adjust the model described here and the experimental data, we have used the ``Particle Swarm Optimization'' method~\cite{i53,i54}. By using this method the real and imaginary part of the impedance are simultaneous adjusted with the experimental data, in particular for this case the adjusted ${\cal{R}}^2$~\cite{i55,i56} points  out that the model account for about $99.9\%$ of the observed variance in the experimental data.

\section{Concluding remarks}

The electrical impedance response of a system governed in the bulk by a fractional diffusion equation of distributed order in presence of a reaction term and subjected to a general boundary condition was theoretically investigated. This boundary condition was built in such a way to cover a broad scenario of physical situations, ranging from the ones characterized by charge transfer till to selective adsorption, including specific ion adsorption described by the generalized Chang-Jaff\'e boundary conditions. In this enlarged scenario, an exact expression for the electrical impedance in the small a.c. signal limit was found. Analysis of the predictions from this exact, and general,  expression allows one to conclude that in the low frequency domain,  in absence of the reaction term, the system is governed by $\overline{k}_{e}\neq 0$  (different Ohmic parameters) even if the adsorption-desorption process is present. This feature is clearly illustrated in Figs.~\ref{Figura1}a and~\ref{Figura5}b for the real and imaginary parts of the impedance, and can be analytically determined for the cases characterized by $\Upsilon_{a}(i\omega)$ $=$ $\kappa_{a}\tau(i\omega\tau)^{1-\delta}$, when  $\omega\rightarrow 0$, with  $0<\delta< 1$. The influence of the adsorption may be relevant in the low frequency limit,  when Chang-Jaffé-like boundary conditions are considered. This implies that   $\Upsilon_{a}(i\omega)$ $=$ $\kappa_{CJ}\tau$, for  $\omega\rightarrow 0$. Figures~\ref{Figura3}a and ~\ref{Figura3}b exhibit results for a suitable combination of the situations illustrated in Fig.~\ref{Figura1},  with $\Upsilon_{a}(i\omega)$ $=$ $\kappa_{CJ}\tau$ $+$ $\overline{\Upsilon}_{a}(i\omega)$, for  $\gamma\neq 1$. The results show that, in the limit of low frequency, the surface effects may govern the dynamic behavior of the mobile charges. The influence of a reaction term on the IS response was analyzed in details. Likewise, a comparison between the theoretical predictions and some experimental data was carried out,  in order to show how the formalism developed here can be used to describe (with good agreement) the IS response of a finite-length situations in a typical electrolytic cell.  In particular, in the low frequency limit, it was shown that the electrical response is truly affected by the choice of the reaction term governing the bulk behavior of the mobile charges. These results reinforce the persuasion that the general framework presented here may be helpful to face real problems connected with the IS response of electrolytic cells. In this new formalism,  it is also possible to describe relevant and complex scenarios because, besides containing as particular cases many others useful models already proposed,  it accounts,  in a synthetic and mathematically unified way,  also for the possible presence of different regimes (usual or anomalous) for the diffusion of the mobile charges. In addition, the present framework can be connected with other formalisms as the ones based on equivalent circuits with constant-phase elements (CPE) as analytically demonstrated in Ref.~\cite{i52} with a suitable choice of the kernels present in the boundary conditions. In other words, the model proposed here contains, as particular cases that depend on the specific situation considered, equivalent circuits with CPE. Thus, a formulation of this kind provides  a simple interpretation of these constant-phase elements in terms of a continuum PNPA description, i.e., an approach that uses fractional calculus with general boundary conditions.

\section*{Acknowledgement}
This work was partially supported by the National Institutes of Science and Technology (INCT-CNPq) of Complex Systems (E.K.L.) and Complex Fluids (L. R. E.), and by the Brazilian Agency, Capes (F. R. G. B. Silva).

\end{document}